\documentclass[prb,a4paper,twocolumn,floatfix,superscriptaddress,amsmath,amsfonts,amssymb,citeautoscript,preprintnumbers,longbibliography]{revtex4-2}
\pdfoutput=1 \widowpenalty10000 \clubpenalty10000
\usepackage[T1]{fontenc}\usepackage[latin1]{inputenc}
\usepackage{dcolumn,graphicx,color,booktabs,microtype,afterpage,upgreek,fixmath,relsize}
\usepackage[slantedGreek]{mathptmx}
\usepackage{pzccal}

\newcount\hh \newcount\mm
\hh=\time \divide\hh by 60
\mm=\hh \multiply\mm by 60 \mm=-\mm
\advance\mm by \time
\def\now{\number\hh:\ifnum\mm<10{}0\fi\number\mm}

\usepackage{hyperref}
\hypersetup{colorlinks, plainpages=false, linkcolor=blue, urlcolor=blue, citecolor=blue, pdfpagemode=UseNone, pdfstartview=FitBH}

\newcommand{\pmmp}{^\text{\raisebox{-0.5pt}{\textpm}}{\scriptstyle\!\!\!\text{\raisebox{0.5pt}{$\diagup$}}\!\!\!}_\text{\rotatebox[origin=c]{180}{\raisebox{-0.5pt}{\textpm}}}\,}
\newcommand{\textmp}{\rotatebox[origin=c]{180}{\textpm}}
\newcommand{\hamcal}{\kern3pt\hat{\kern-3pt\mathcal{H}}\kern-1pt}

\begin{document}


\title{Local origin of the strong field-space anisotropy in the magnetic phase diagrams of Ce$_{\text{1}-\text{\it x}}$La$_\text{\it x}$B$_\text{6}$ measured in a rotating magnetic field\smallskip\smallskip}

\author{D.~S. Inosov}\email[Corresponding author:~]{dmytro.inosov@tu-dresden.de}
\affiliation{\mbox{Institut f\"ur Festk\"orper- und Materialphysik, Technische Universit\"at Dresden, H\"ackelstra{\ss}e 3, 01069 Dresden, Germany}}
\affiliation{\mbox{W\"urzburg-Dresden Cluster of Excellence on Complexity and Topology in Quantum Matter\,---\,\textit{ct.qmat}, TU~Dresden, 01069 Dresden, Germany}}

\author{S.~Avdoshenko}
\affiliation{\mbox{Leibniz-Institut f\"ur Festk\"orper- und Werkstoffforschung (IFW) Dresden, Helmholtzstra{\ss}e 20, 01069 Dresden, Germany}}

\author{P.~Y.~Portnichenko}
\affiliation{\mbox{Institut f\"ur Festk\"orper- und Materialphysik, Technische Universit\"at Dresden, H\"ackelstra{\ss}e 3, 01069 Dresden, Germany}}

\author{Eun Sang Choi}
\affiliation{\mbox{National High Magnetic Field Laboratory, Florida State University, Tallahassee, Florida 32310-3706, USA}}

%

\author{A.~Schneidewind}
\affiliation{\mbox{J\"ulich Center for Neutron Science at MLZ, Forschungszentrum J\"ulich GmbH, Lichtenbergstra{\ss}e 1, 85748 Garching, Germany}}

\author{J.-M. Mignot}
\affiliation{\mbox{Laboratoire L\'eon Brillouin, CEA-CNRS, CEA/Saclay, 91191 Gif sur Yvette, France}}

\author{M.~Nikolo}\email[Corresponding author:~\vspace*{-2pt}]{martin.nikolo@slu.edu}
\affiliation{\mbox{Department of Physics, Saint Louis University, St. Louis, Missouri 63103, USA}}


\begin{abstract}
Cubic $f\!$-electron compounds commonly exhibit highly anisotropic magnetic phase diagrams consisting of multiple long-range-ordered phases. Field-driven metamagnetic transitions between them may depend not only on the magnitude, but also on the direction of the applied magnetic field. Examples of such behavior are plentiful among rare-earth borides, such as $R$B$_6$ or $R$B$_{12}$ ($R$~=~rare earth). In this work, for example, we use torque magnetometry to measure anisotropic field-angular phase diagrams of La-doped cerium hexaborides, Ce$_{1-x}$La$_x$B$_6$ ($x=0$, 0.18, 0.28, 0.5). One expects that field-directional anisotropy of phase transitions must be impossible to understand without knowing the magnetic structures of the corresponding competing phases and being able to evaluate their precise thermodynamic energy balance. However, this task is usually beyond the reach of available theoretical approaches, because the ordered phases can be noncollinear, possess large magnetic unit cells, involve higher-order multipoles of 4$f$ ions rather than simple dipoles, or just lack sufficient microscopic characterization. Here we demonstrate that the anisotropy under field rotation can be qualitatively understood on a much more basic level of theory, just by considering the crystal-electric-field scheme of a pair of rare-earth ions in the lattice, coupled by a single nearest-neighbor exchange interaction. Transitions between different crystal-field ground states, calculated using this minimal model for the parent compound CeB$_6$, possess field-directional anisotropy that strikingly resembles the experimental phase diagrams. This implies that the anisotropy of phase transitions is of local origin and is easier to describe than the ordered phases themselves.
\end{abstract}

\maketitle

\section{\label{Sec:Introduction}Introduction}\vspace{-2pt}

\subsection{\label{SubSec:AnisotropyCubic}Magnetic anisotropy in cubic systems}\vspace{-4pt}

It is often believed that magnetocrystalline anisotropy in cubic magnetic systems must usually be small. This may be true for systems with vanishingly weak spin-orbit coupling, because the contribution from the spin-spin part of the electron interactions is known to average out in collinear cubic magnets \cite{Vleck37, Jansen88}. It is nevertheless possible to observe significant field-angular anisotropy of magnetic and transport properties in certain transition-metal compounds such as the cubic perovskite SrCoO$_3$, where they have been attributed to orbital fluctuations and to an anisotropic scattering rate of conduction electrons in the ferromagnetic state \cite{LongKaneko11}, and even in the spin-1/2 cubic antiferromagnet Cu$_3$TeO$_6$ \cite{HerakBerger05, Herak11}.

Strong magnetic anisotropy is especially common in rare-earth (RE) compounds, where it is due to the effect of crystalline electric fields (CEF) on the $4f$-electron wave functions. For example, it is very pronounced in $R$Fe$_2$ and $R$Co$_2$ Laves phases \cite{ClarkBelson72, SamataFujiwara99} ($R$~=~rare earth), where it could be satisfactorily explained by means of single-ion crystal-field theory \cite{Buschow77}. From more recent examples, strong anisotropy in the field-angular dependence of magnetoresistance, measured upon continuously rotating the direction of an external magnetic field $\mathbf{B}$, was reported for the half-Heusler compounds TbPtBi and HoPtBi \cite{Roncaioli19, PavlosiukFalat20} and for several RE dodecaborides \cite{SluchankoKhoroshilov18, KhoroshilovKrasnorussky19, SluchankoKhoroshilov20, KrasikovBogach20, KrasikovGlushkov20, KrasikovAzarevich20, AzarevichBogach20}, most prominently pure and doped TmB$_{12}$ and ErB$_{12}$. In TmB$_{12}$ and its doped derivatives Tm$_{1-x}$Lu$_{x}$B$_{12}$ and Tm$_{1-x}$Yb$_{x}$B$_{12}$, the field-angular magnetic phase diagrams in the $\mathbf{B}\perp\langle110\rangle$ plane, suggested by the magnetotransport measurements, resemble a Maltese cross \cite{KhoroshilovKrasnorussky19, SluchankoKhoroshilov20, KrasikovBogach20, KrasikovGlushkov20, KrasikovAzarevich20, AzarevichBogach20} with very sharp straight transition lines, separating different antiferromagnetic (AFM) phases that exist between 0 and 3~T in sectors around the [001], [111], and [110] field directions. Doping with nonmagnetic lutetium tends to enhance this anisotropic behavior.

Still, even in systems with strong spin-orbit coupling (such as heavy-fermion systems with $f$ electrons), experimentalists commonly do not anticipate either a strong dependence of the system's magnetic properties on the direction of the applied magnetic field or any qualitative change in the magnetic ground-state configuration upon field rotation. Observations of strong field-directional anisotropy in magnetization, specific heat, or magnetotransport data are sometimes interpreted as evidence for spontaneously broken cubic symmetry (e.g., due to the formation of charge stripes \cite{BolotinaDudka18, SluchankoKhoroshilov18, KhoroshilovKrasnorussky19, SluchankoKhoroshilov20, KrasikovBogach20, KrasikovAzarevich20, AzarevichBogach20, KrasikovGlushkov20} or electron-nematic instabilities \cite{SemenoGilmanov16, DemishevKrasnorussky17}) even without quantitative estimates of the expected magnitude of the same effect in the original cubic symmetry of the lattice. In the present work, we provide such an estimate by calculating the field-angular anisotropy of the CEF ground state for a minimal two-site model consisting of a pair of RE ions in a cubic crystal field, coupled by a single exchange interaction. The resulting field-angular phase diagrams are qualitatively similar to those obtained experimentally and display equally strong anisotropy. This suggests that the anisotropy of phase boundaries separating long-range-ordered magnetic phases in CeB$_6$ are of much more fundamental origin than these phases themselves and appears even within a purely local model that does not break the cubic lattice symmetry.

\vspace{-3pt}\subsection{\label{SubSec:IntroCeB6}Magnetic phase diagram of Ce$_{\text{1-\textit{x}}}$La$_{\text{\rule{0pt}{\fontcharht\font`1}\textit{x}}}$B$_\text{6}$}\vspace{-4pt}

The heavy-fermion compound CeB$_\text{6}$ is well suited for our purpose because its magnetic phase diagram contains several long-range-ordered phases separated by field-driven magnetic phase transitions that are very well studied (see Ref.~\citenum{CameronFriemel16} and references therein). In zero magnetic field, the ground state is a double-$\mathbf{q}$ AFM phase~III with a noncollinear spin arrangement characterized by a pair of propagation vectors, $\mathbf{q}_1=\left(\frac{1}{4}\frac{1}{4}0\right)$ and $\mathbf{q}_2=\left(\frac{1}{4}\frac{1}{4}\frac{1}{2}\right)$; in moderate fields (1--2~T, depending on the field direction) it changes to a single-$\mathbf{q}$ collinear AFM phase~III$^\prime$ \cite{HornSteglich81, EffantinRossat-Mignod85, KusunoseKuramoto01, ZaharkoFischer03, KunimoriKotani11}; and at even higher fields the order parameter of the ground state changes one more time to phase~II that represents a combination of primary antiferroquadrupolar (AFQ) order with the propagation vector $\mathbf{q_\text{AFQ}}=\left(\frac{1}{2}\frac{1}{2}\frac{1}{2}\right)$, also referred to as \emph{orbital antiferromagnetic} in earlier literature \cite{Ohkawa85}, and a secondary field-induced dipolar-octupolar order with the same wave vector \cite{Rossat-Mignod87, ErkelensRegnault87, ShiinaShiba97, SakaiShiina97, ShiinaSakai98, SeraKobayashi99, SeraIchikawa01, MatsumuraYonemura09, MatsumuraYonemura12}. Substitution of nonmagnetic La for Ce in the Ce$_{1-x}$La$_{x}$B$_6$ solid solutions \cite{CameronFriemel16} stabilizes another multipolar phase~IV, presumably antiferrooctupolar (AFO) in character \cite{TayamaSakakibara97, HiroiKobayashi97, HiroiKobayashi98, KobayashiSera00, KobayashiYoshino03, KuwaharaIwasa07, LoveseyFernandez-Rodriguez07, KuwaharaIwasa09, MatsumuraMichimura14, NikitinPortnichenko18, SeraKunimori18, HanzawaYamada19}.

It has long been known that the magnetic phase diagrams of Ce$_{1-x}$La$_{x}$B$_6$ depend on the direction of magnetic field. For example, the transition to phase~II in pure CeB$_{6}$ occurs at 2.3~T for $\mathbf{B} \parallel \langle001\rangle$ but already at 1.4~T for $\mathbf{B} \parallel \langle110\rangle$ \cite{HornSteglich81}. The intermediate phase III$^\prime$ extends from 1.0 to 1.65~T for $\mathbf{B} \parallel \langle111\rangle$, but nearly disappears for $\mathbf{B} \parallel \langle110\rangle$~\cite{KunimoriKotani11}. The anisotropy becomes even stronger upon La substitution, as phase~III extends up to 4~T in Ce$_{0.8}$La$_{0.2}$B$_6$~\cite{KobayashiYoshino03} and up to 4.5~T in Ce$_{0.5}$La$_{0.5}$B$_6$~\cite{TayamaSakakibara97} for $\mathbf{B} \parallel \langle001\rangle$ but remains nearly unchanged for $\mathbf{B} \parallel \langle111\rangle$ \cite{KomatsubaraSato83}. The available published data are abundant, but they are scattered over many publications, originate from measurements at different temperatures using different methods, and are mostly restricted to high-symmetry field directions with the single exception of the field-angular dependence of magnetization in pure CeB$_6$ measured at 1.4~K by Kunimori \textit{et al.}~\cite{KunimoriKotani11}. Therefore, in this paper we use torque magnetometry to measure the field-angular phase diagrams of Ce$_{1-x}$La$_x$B$_6$ ($x=0$, 0.18, 0.28, 0.50) systematically under continuous field rotation and at a much lower temperature of 20~mK. To identify the magnetic phases, we compare the results with elastic neutron scattering data and with a panoply of previously published data.

\vspace{-1pt}\section{\label{Sec:Experiment}Experimental results}\vspace{-2pt}

\subsection{\label{SubSec:NeutronScattering}Elastic neutron scattering}\vspace{-4pt}

\begin{figure}[b]
\centerline{\includegraphics[width=0.815\columnwidth]{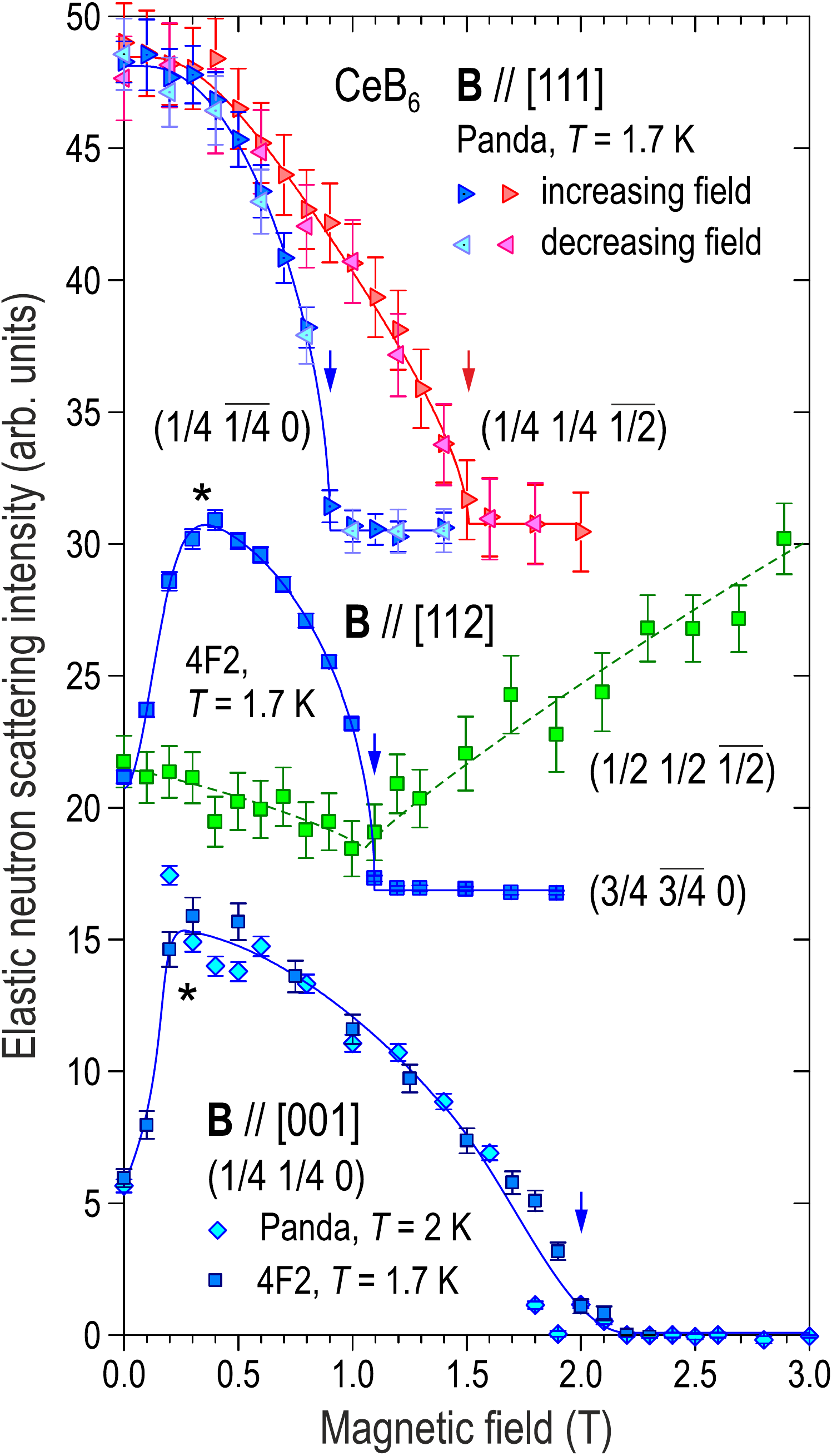}}
\caption{\label{Fig:NeutronScattering} Magnetic field dependence of the elastic neutron scattering intensity in pure CeB$_{6}$ at $\mathbf{q}_1=\left\{\frac{1}{4}\frac{1}{4}0\right\}$ or $\left\{\frac{3}{4}\frac{3}{4}0\right\}$, $\mathbf{q}_2=\left\{\frac{1}{4}\frac{1}{4}\frac{1}{2}\right\}$, and $\mathbf{q}_\text{AFQ}=\left\{\frac{1}{2}\frac{1}{2}\frac{1}{2}\right\}$ for three different field directions: $\mathbf{B}\parallel[111]$, $[112]$, $[001]$. The data were measured in different experiments with different crystal orientations, therefore intensities are not comparable. The datasets are shifted vertically for clarity. Phase transitions are marked with arrows, the domain selection transition is indicated with asterisks.\vspace{-3pt}}
\end{figure}

Neutron scattering is one of the most direct ways to reveal the character and microscopic structure of a magnetic phase. To simplify the correct identification of phase transitions in the following, we start with presenting elastic neutron scattering data for pure CeB$_6$. In Fig.~\ref{Fig:NeutronScattering}, we show the field dependences of magnetic Bragg intensity in CeB$_6$, which were measured at the ordering vectors of phases II, III, and III$^\prime$ for several high-symmetry directions of the magnetic field, $\mathbf{B}\parallel[111]$, $[112]$, and $[001]$. The data for $\mathbf{B}\parallel[110]$ from the same sample can be found elsewhere~\cite{CameronFriemel16}. The measurements were done using cold-neutron triple-axis spectrometers \textsc{Panda} at the FRM-II research reactor of the Maier-Leibnitz Zentrum (MLZ, Garching, Germany) \cite{SchneidewindLink06, SchneidewindCermak15} and \textsc{4F2} at the reactor Orph\'ee of the Laboratoire L\'eon Brillouin (LLB, CEA-Saclay, France), in addition to the spectroscopic measurements described in Ref.~\citenum{PortnichenkoAkbari20}. Energy analysis was used to eliminate the contribution from inelastic scattering, revealing purely elastic Bragg intensity at $\mathbf{q}_1=\left\{\frac{1}{4}\frac{1}{4}0\right\}$ or $\left\{\frac{3}{4}\frac{3}{4}0\right\}$, $\mathbf{q}_2=\left\{\frac{1}{4}\frac{1}{4}\frac{1}{2}\right\}$, and $\mathbf{q}_\text{AFQ}=\left\{\frac{1}{2}\frac{1}{2}\frac{1}{2}\right\}$. These measurements show a factor of 2 anisotropy in the field required to suppress phase~III between the $[111]$ and $[001]$ field directions.

It has to be noted that due to the geometric constraints of the triple-axis experiment, not every ordering vector can be reached for certain field directions. The measurement is restricted to wave vectors in the horizontal scattering plane, while the field direction is vertical in all our measurements \cite{PortnichenkoAkbari20}. Therefore, $\mathbf{q}_1$ can be probed in all four high-symmetry configurations (but not at intermediate angles), $\mathbf{q}_2$ is only accessible for $\mathbf{B}\parallel[111]$ and $[110]$, and $\mathbf{q}_\text{AFQ}$ only for $\mathbf{B}\parallel[110]$ and $[112]$. The neutron data reveal three distinct phase transitions:

\textit{1.~Domain selection.} At low fields, the selection of AFM domains leads to an enhancement (or suppression) of the Bragg intensity at $\mathbf{q}_{1,2}$ for any general field direction except for $\mathbf{B}\parallel[111]$, because this field orientation does not break the equivalency of domains with $\left(\frac{1}{2}\frac{1}{2}0\right)$, $\left(\frac{1}{2}0\frac{1}{2}\right)$, and $\left(0\frac{1}{2}\frac{1}{2}\right)$ propagation vectors. For $\mathbf{B}\parallel[001]$ or $[112]$, domains with the wave vector orthogonal to field are favored, which results in a threefold increase in Bragg intensity at low fields at the expense of the suppressed domains whose ordering vectors lie above or below the scattering plane. The domain-selection transition is complete around 0.3~T (asterisk symbols in Fig.~\ref{Fig:NeutronScattering}). For $\mathbf{B}\parallel[110]$, the Bragg peaks in the scattering plane show opposite behavior and are suppressed by the field \cite{CameronFriemel16}.

\textit{2.~Phase~III--III$^\prime$ transition.} As one can see from the $\mathbf{B}\parallel[111]$ data, the Bragg intensity at $\mathbf{q}_1$ is suppressed before that at $\mathbf{q}_2$. This signifies a transition from the double-$\mathbf{q}$ phase III to the single-$\mathbf{q}$ phase III$^\prime$. As one can see from Fig.~\ref{Fig:PhaseDiagrams}, this intermediate phase is most stable around $\mathbf{B}\parallel[111]$, where it occupies a field range from 1.1 to 1.7~T at $T=20$~mK or from 0.9 to 1.5~T at $T=1.7$~K. The rotation of magnetic-field direction towards $[001]$ stabilizes phase III so that it persists up to 2~T. The field range occupied by phase III$^\prime$ within the $\mathbf{B}\perp[001]$ plane is reduced to approximately 0.2~T (Fig.~\ref{Fig:CeB6_HK0}), which is consistent with earlier magnetization measurements by Kunimori \textit{et al.}~\cite{KunimoriKotani11}.

\textit{3.~The transition to phase~II.} The primary AFQ order parameter in phase~II is ``hidden'' to neutron scattering, because quadrupolar moments are nonmagnetic and have zero scattering cross section. Nevertheless, the onset of AFQ order in phase~II is manifested by the secondary field-induced dipolar-octupolar order parameter in phase~II \cite{MatsumuraYonemura12} that leads to a linear increase in Bragg intensity at $\mathbf{q}_\text{AFQ}$ \cite{Rossat-Mignod87, SeraIchikawa01, CameronFriemel16}. In pure CeB$_6$, this onset coincides with the suppression of phase~III, as one can see in the data for $\mathbf{B}\parallel[112]$ in Fig.~\ref{Fig:NeutronScattering}. Upon La doping, as the region corresponding to phase III$^\prime$ increases, the magnetic intensity at $\mathbf{q}_\text{AFQ}$ acquires two-step behavior \cite{JangPortnichenko17} with an initial intensity onset already in phase III$^\prime$ and a second stronger increase upon the transition from phase III$^\prime$ to phase~II. This implies that AFM and AFQ order parameters coexist within phase III$^\prime$, in contrast with the purely dipolar phase~III \cite{JangPortnichenko17}. This also gives us a way to measure both phase transitions at the same wave vector, which is especially useful for those field directions where $\mathbf{q}_1$ and $\mathbf{q}_2$ cannot be reached simultaneously. It is not clear so far whether the same two-step behavior can be seen in CeB$_6$, e.g., for $\mathbf{B}\parallel[111]$.

Apart from neutron scattering, phase transitions in Ce$_{1-x}$La$_{x}$B$_6$ can be sensitively probed by resonant x-ray diffraction, and the associated structural distortions by high-resolution x-ray diffraction. For instance, the first-order transition \cite{TayamaSakakibara97} from phase IV to phase III in Ce$_{0.7}$La$_{0.3}$B$_6$ has been determined very accurately for $\mathbf{B}\parallel[110]$ by the rhombohedral splitting of the $(444)$ structural Bragg peak in synchrotron x-ray diffraction \cite{InamiMichimura14}. The same transition was also observed by elastic neutron scattering \cite{JangPortnichenko17} and by resonant x-ray diffraction measured with different polarizations \cite{MatsumuraMichimura14} as sharp anomalies in the field dependence of the magnetic Bragg intensity at $\left(\frac{1}{2}\frac{1}{2}\frac{1}{2}\right)$ and $\left(\frac{3}{2}\frac{3}{2}\frac{1}{2}\right)$, respectively. Finally, field-driven phase transitions in Ce$_{1-x}$La$_{x}$B$_6$ were also determined accurately at subkelvin temperatures using specific heat \cite{NakamuraEndo03, JangPortnichenko17, GrunerKim19}, resistivity \cite{HiroiSera97, HiroiKobayashi97, HiroiKobayashi98, KobayashiSera00, NakamuraSato03}, magnetization \cite{WinzerFelsch78, KomatsubaraSato83, TayamaSakakibara97, KunimoriTanida10}, and ultrasonic \cite{NakamuraGoto95, KobayashiYoshino03, AkatsuGoto04, SuzukiNakamura05} measurements. While there is a good general agreement among all these results, the measurements are mostly restricted to a few high-symmetry field directions, so in order to reconstruct the magnetic phase diagrams as a function of field angle, we proceed with presenting datasets taken under continuous field rotation at $T=20$~mK.

\vspace{-4pt}\subsection{\label{SubSec:Magnetization}Torque magnetometry}\vspace{-4pt}

\begin{figure}[b]
\includegraphics[width=\columnwidth]{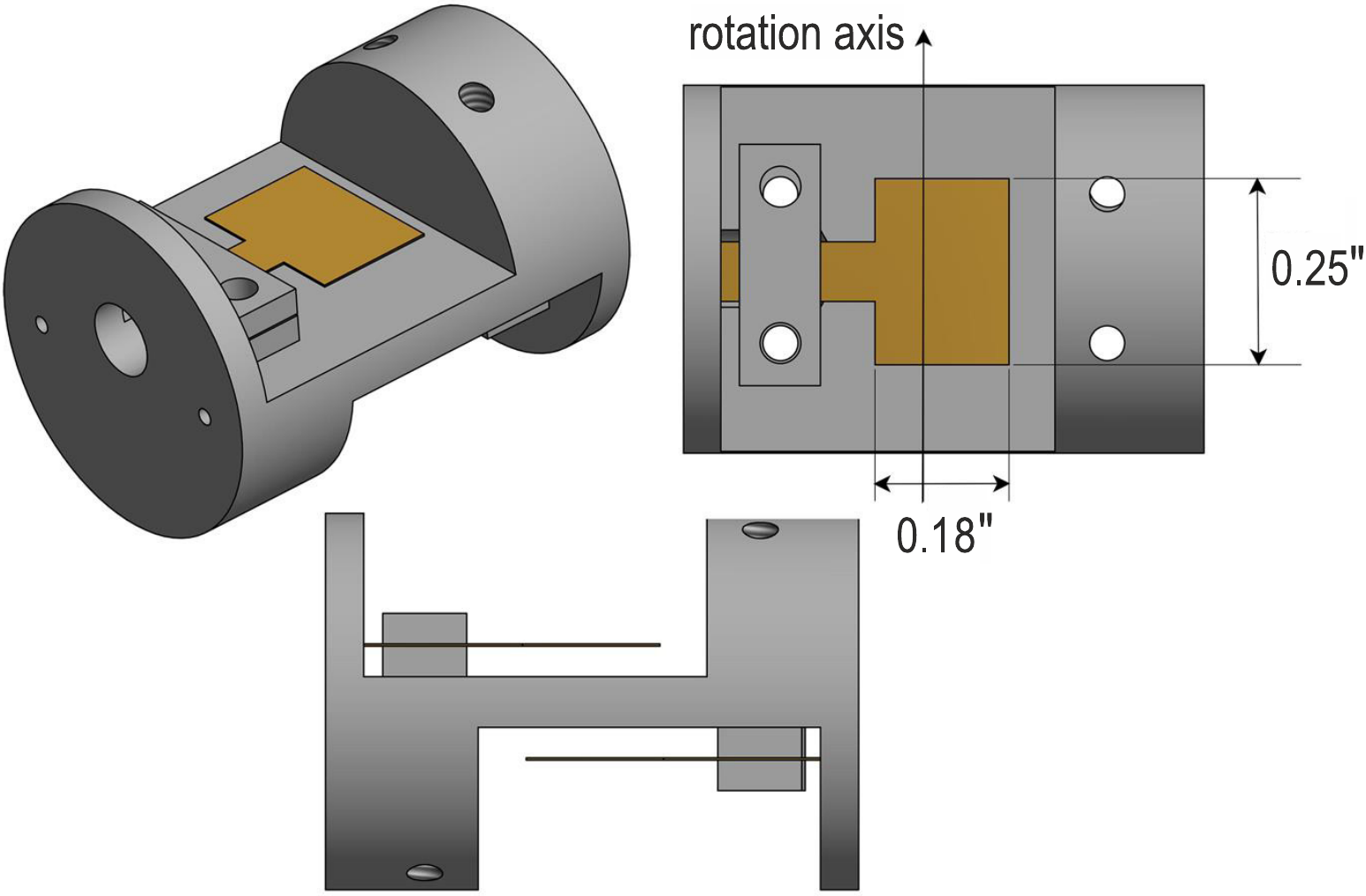}
\caption{Design of the sample holder for cantilever magnetometry at NHMFL with a pair of cantilevers on opposite sides.\vspace*{-4pt}}
\label{Fig:TorqueHolder}
\end{figure}

\begin{figure}[t]
\includegraphics[width=0.65\columnwidth]{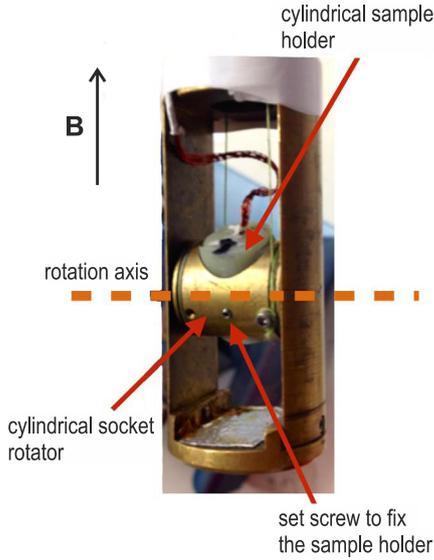}
\caption{Sample holder is mounted in a cylindrical socket rotator \protect\cite{SCM1_PDF}. The magnetic field direction is vertical.}
\label{Fig:TorqueRotator}
\end{figure}

Torque magnetometry is a highly sensitive and powerful method used to determine and analyze the magnetic anisotropy in magnetic materials. The general principle behind torque magnetometry is mounting of a sample on a flexible cantilever and measuring its elastic deflection caused by the magnetic torque $\mathbold{\tau} = \mathbf{M}(\mathbf{B}) \times \mathbf{B}$ that is exerted on the sample with magnetization $\mathbf{M}$ by an external magnetic field $\mathbf{B}$. This method therefore probes the difference in direction between the magnetization and the applied field, which results from magnetocrystalline anisotropy. The cantilever is mounted on one end like a diving board, and its deflection (proportional to the magnetic torque) is measured capacitively with high accuracy. This technique was developed by J.~S. Brooks \textit{et al.} at Francis Bitter Magnet Laboratory--MIT in 1987 \cite{BrooksNaughton87} and has been adopted by a number of major laboratories \cite{WiegersvanSteenbergen98, MatthewsUsher04, OhmichiYoshida03, ZhouChoi10, AdhikariKaundal12, ModicBachmann18, MumfordPaul20}, including the National High Magnetic Field Laboratory (NHMFL), Tallahassee, FL, where D.~Hall documented the method in detail \cite{Hall99}.

Two such cantilevers made of beryllium copper (a nonmagnetic alloy with good elastic properties) with mounted samples are placed in a $\varnothing$12.6~mm cylindrical sample holder shown in Fig.~\ref{Fig:TorqueHolder}, which is designed so one can measure two different samples simultaneously \cite{SCM1_PDF}. Measurements of the cantilever capacitance (of the order of 1~pF) are made with an Andeen-Hagerling 2700A capacitance bridge. The holder is inserted to a cylindrical socket rotator \cite{PalmMurphy99} (Fig.~\ref{Fig:TorqueRotator}), which is then mounted inside a top-loading dilution refrigerator with a permanently installed 18/20~T superconducting magnet at NHMFL, also known as SCM1. It has a base temperature of 20~mK and 400~$\upmu$W of cooling power at 100~mK.

In the present work, single-crystal samples of Ce$_{1-x}$La$_x$B$_6$ with $x=0$ (5.5~mg), $x=0.18$ (3.6~mg), $x=0.28$ (5.0~mg), and $x=0.5$ (5.0~mg) have been investigated. Magnetic torque was measured in magnetic fields from 0 to 18~T, yet for the purpose of this paper we are only interested in data below 5~T, because the data at higher fields turned out featureless, which is consistent with the absence of phase transitions in our samples in this field range. We made several trial runs to find the optimal thickness of the beryllium copper cantilevers. The parent CeB$_6$ compound was measured with a 50-$\upmu$m cantilever, whereas the La-doped samples were initially measured with 25-$\upmu$m cantilevers, but these thinner cantilevers were very sensitive and occasionally touched the base plate already at intermediate magnetic fields. We replaced them with thicker ones and in the end, the cantilevers used ranged in thickness from 25 to 125 $\upmu$m. We also experimented with gluing two thinner cantilevers together.

We used a Laue camera and a single-crystal diffractometer to precisely align either [001] or $[1\overline{1}0]$ sample direction along the rotation axis on the $0.18^{\prime\prime} \times 0.25^{\prime\prime}$ cantilever paddle (see Fig.~\ref{Fig:TorqueHolder}). The samples were attached to the cantilever paddle with either $\langle100\rangle$ or $\langle110\rangle$ surfaces using either GE varnish or \textsc{Loctite}\circledR\ STYCAST 2850FT. The rotation angle was controlled by a stepper motor with a resolution of 110 steps per 1$^\circ$ angle, giving the rotator probe about 0.02$^\circ$ of resolution. The measurement uncertainties came from how well the cantilevers were aligned to the holder and how well the holder was mounted on the probe. To minimize those, we had a Hall probe sensor mounted on the holder, which gave us orientation of the holder plate. The sample's response was usually the best way to re-confirm its orientation, because high-symmetry directions in the crystal correspond to the orientations where the magnetic torque vanishes. The pair of cantilevers on opposite sides of the sample holder allowed us to perform measurements in the $[1\overline{1}0]$ and [001] planes of field rotation by measuring two samples mounted in corresponding orientations simultaneously or to measure two different samples at a time.

For most of our runs, the samples were rotated at zero field or at constant fields at the base temperature of 20~mK. The Hall signal was recorded to give a rough idea of the cantilever angle with respect to the field. Therefore, our data files typically recorded torque (arbitrary units), Hall voltage (V), magnetic field (T), temperature (K), and orientation as a four-digit readout from the servomotor, which was later converted to an angle (degrees). Field sweeps were performed with both increasing and decreasing field, which resulted in practically identical datasets within experimental uncertainties. Selected angular sweeps at fixed fields were also done to verify the crystal orientation by the symmetry of the signal. The field sweeps from 0 to 18~T would take between 3~h at 20~mK and 1~h at $T>0.3$~K. The sweep rates were typically 0.2 or 0.3~T/min. Most angular sweeps were run in steps of $2.5^\circ$ to cover an angular range of at least $120^\circ$, i.e., broader than the irreducible part of the field-angular phase diagram.

\begin{figure}[b]
\includegraphics[width=\columnwidth]{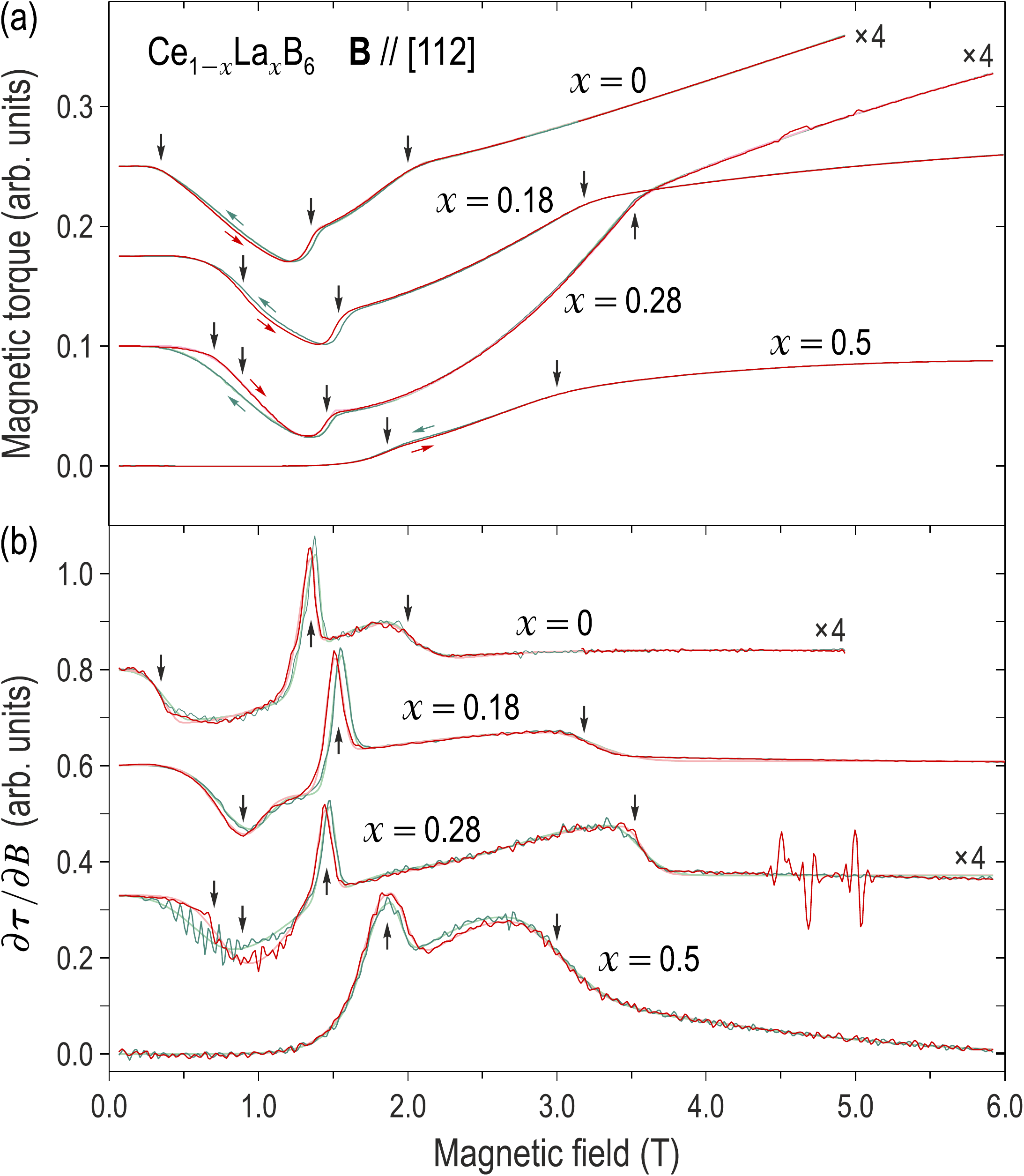}
\caption{(a)~Magnetic-field dependence of the torque, $\mathbold{\tau}(B)$, for the magnetic field direction $\mathbf{B}\parallel[112]$, measured on four samples of Ce$_{1-x}$La$_x$B$_6$ ($x=0$, 0.18, 0.28, 0.50) during upward and downward field sweeps, as indicated with arrows. (b)~Magnetic-field derivative, $\partial\mathbold{\tau}/\partial B$, of the data in panel (a). The curves in both panels are shifted vertically for clarity. Smooth pale lines in the background are empirical fits to the data. Vertical black arrows mark the estimated phase transition fields.\vspace{-2pt}}
\label{Fig:Torque112}
\end{figure}

\begin{figure*}
\includegraphics[width=\textwidth]{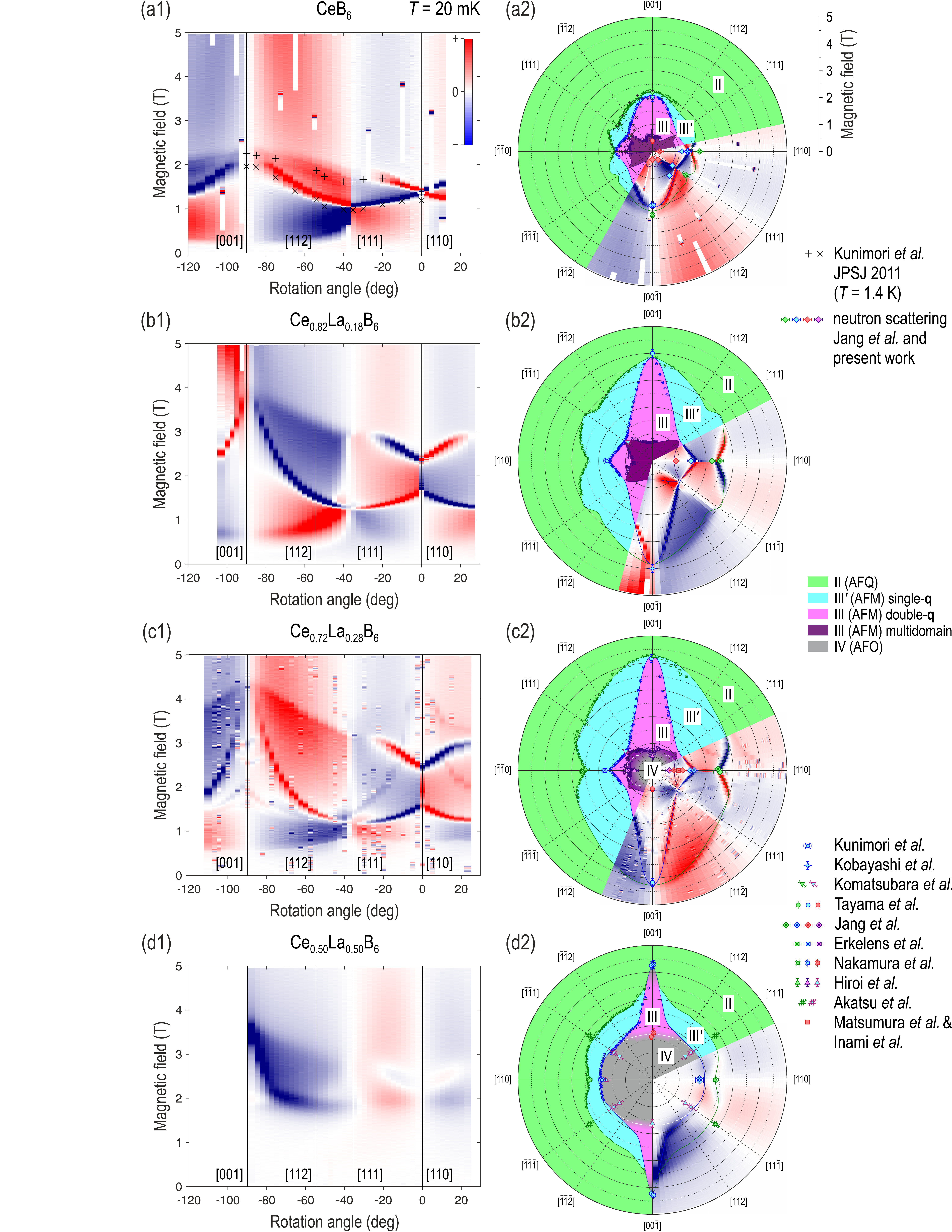}
\caption{Field-angular magnetic phase diagrams of Ce$_{1-x}$La$_x$B$_6$ (from top to bottom: $x=0$, 0.18, 0.28, 0.50) as measured by torque magnetometry at $T=20$~mK. The field is rotated in the $[1\overline{1}0]$ plane. The field derivative of the magnetic torque, $\partial\mathbold{\tau}/\partial B$, measured in field sweeps upon increasing the field, is plotted for every sample on the left as a color map. The corresponding field-angular phase diagrams in polar coordinates are presented on the right. The angular dependence for CeB$_6$ at $T = 1.4$~K from Ref.~\citenum{KunimoriKotani11} and the data for high-symmetry directions from our neutron-scattering measurements (Refs.~\citenum{CameronFriemel16}, \citenum{JangPortnichenko17}, and present work) as well as from Refs.~\citenum{KomatsubaraSato83, ErkelensRegnault87, NakamuraGoto95, TayamaSakakibara97, HiroiKobayashi97, HiroiKobayashi98, NakamuraSato03, KobayashiYoshino03, AkatsuGoto04, KunimoriTanida10, MatsumuraMichimura14, InamiMichimura14, CameronFriemel16} are overlaid for comparison.}
\label{Fig:PhaseDiagrams}
\end{figure*}

\begin{figure}[t]
\centerline{\includegraphics[width=\columnwidth]{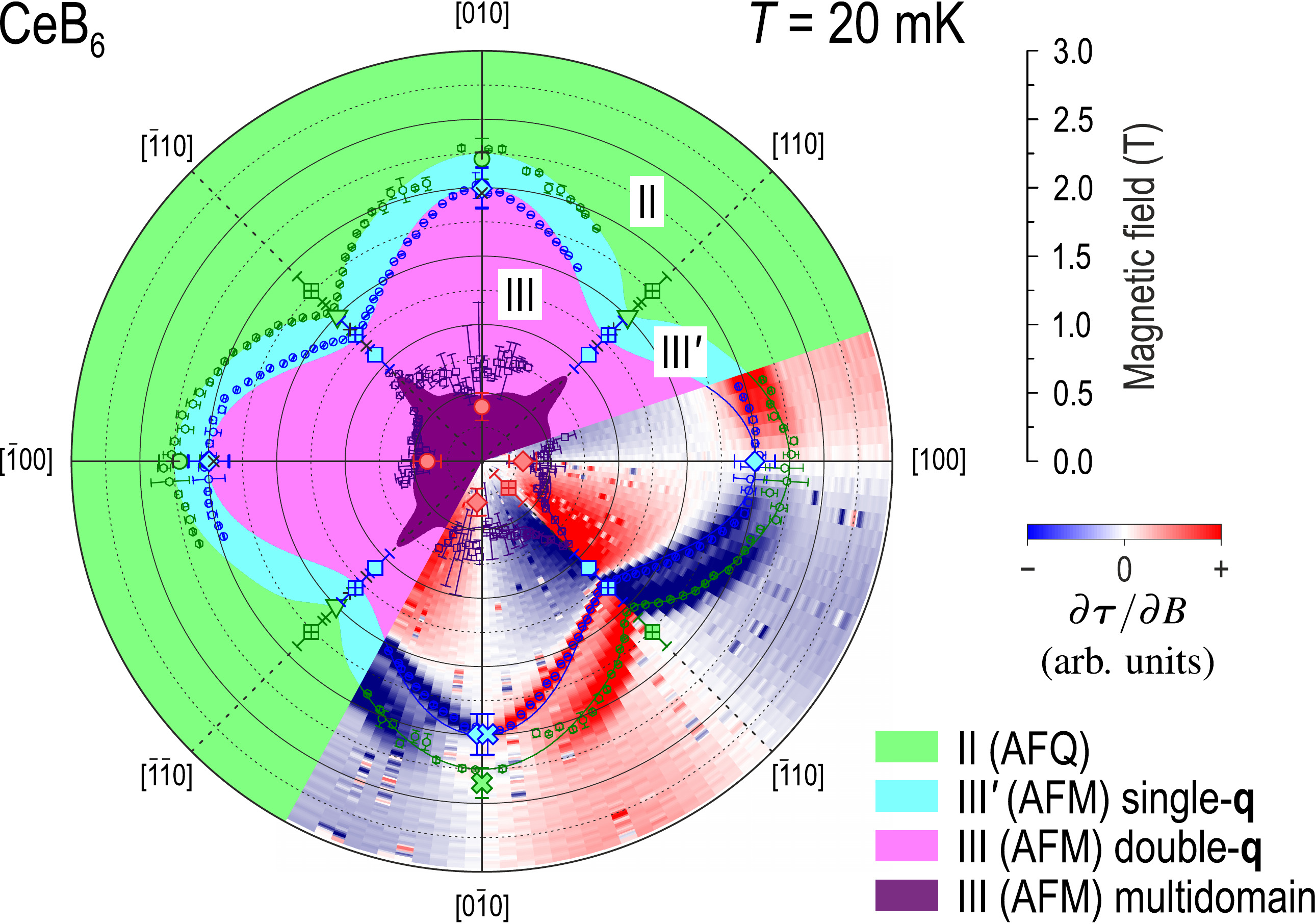}}
\caption{\label{Fig:CeB6_HK0} Polar plot of the field-angular magnetic phase diagram of CeB$_6$ measured at $T=20$~mK in the $C_4$-symmetric field plane, $\mathbf{B} \perp [001]$, using torque magnetometry. The color map shows the field derivative of the magnetic torque, $\partial\mathbold{\tau}/\partial B$, measured in field sweeps upon increasing the field. The data points show fitting results both for the upward (bottom-right sector) and downward (top-left sector) field sweeps, which are essentially identical within the experimental error. The data from our neutron-scattering measurements and from Refs.~\citenum{KomatsubaraSato83, NakamuraGoto95, TayamaSakakibara97, KunimoriKotani11, CameronFriemel16, JangPortnichenko17} are overlaid for comparison (for the meaning of the symbols, see Fig.~\ref{Fig:PhaseDiagrams}).}
\end{figure}

Both the magnetic torque $\tau=\partial F/\partial \theta$ and the magnetization $M=-\partial F/\partial B$ are thermodynamic potentials represented by the first derivatives of the free energy $F(\mathbf{B})$ \cite{ModicBachmann18}. Therefore, the magnetic-field derivative of the torque is equivalent to the angle derivative of the magnetization:
\begin{equation}\label{Eq:Derivatives}
\frac{\partial\mathbold{\tau}(B,\theta)}{\partial B} = \frac{\partial^2 F}{\partial B\partial\theta} = \frac{\partial M(B,\theta)}{\partial\theta}.
\end{equation}
The former quantity can be easily obtained experimentally by differentiating the torque measured during continuous field sweeps, as illustrated in Fig.~\ref{Fig:Torque112}. It can be then compared with the latter quantity, which we calculate in \textit{ab initio} theoretical models as discussed in Sec.~\ref{Sec:Theory}.

We demonstrate our procedure by presenting the data measured for a given field direction, $\mathbf{B}\parallel[112]$. The field dependence of the measured torque and its field derivative for all four samples are shown in Figs.~\ref{Fig:Torque112}\,(a) and \ref{Fig:Torque112}\,(b), respectively. The upward and downward field sweeps show only minor hysteresis effects at some of the field-driven phase transitions, therefore in the following we restrict our consideration to upward field sweeps only. The approximate phase transition fields are indicated with vertical arrows. There is some uncertainty in deciding on the formal criterion for the transition field, which appears to be the main reason for minor discrepancies in the published results obtained by different methods. In our case, step-like transitions (e.g., transition from phase III to phase III$^\prime$) are best defined by the maximum in $\partial\mathbold{\tau}/\partial B$, whereas kink-like transitions (e.g., the transition to phase~II) coincide with the maximal curvature of the $\mathbold{\tau}(B)$ curve or the steepest slope in $\partial\mathbold{\tau}/\partial B$. For some of the transitions, this choice can be ambiguous, especially for closely spaced or broadened transitions. To automate the extraction of phase transition fields from the data, we approximated the measured curves with empirical fitting functions that are shown in Fig.~\ref{Fig:Torque112} with smooth pale-colored lines in the background.

The field-angular phase diagrams that resulted from our torque magnetometry measurements with the field rotated in the $\mathbf{B}\perp[1\overline{1}0]$ plane are shown in Fig.~\ref{Fig:PhaseDiagrams} as color maps of the field derivative of the magnetic torque, $\partial\mathbold{\tau}/\partial B$. The fitted phase transitions and the reference data from our neutron scattering measurements and from earlier publications are shown as data points. In addition, Fig.~\ref{Fig:CeB6_HK0} also shows a field-angular map for CeB$_6$ in the $\mathbf{B}\perp[001]$ plane. One can see that the magnetic torque follows $C_4$ symmetry in the $ab$ plane within the uncertainty of the measurements, as expected in the cubic symmetry. There are known examples among cubic systems where magnetic or transport properties violate the lattice symmetry, for instance due to an anisotropic distribution of lattice defects during crystallization \cite{RosencwaigTabor71} or weak structural distortions \cite{BolotinaDudka18, SluchankoBogach18}. Apparently, CeB$_6$ does not suffer from such complications, so that the cubic symmetry of the magnetization holds to a good approximation.

The AFM phases III and III$^\prime$ are the most anisotropic. Phase~III remains most stable for $\mathbf{B}\parallel[001]$, where it gets additionally stabilized by La substitution, whereas phase III$^\prime$ has the largest stability range for $\mathbf{B}\parallel[111]$. Phase~II is most stable (has the lowest onset field) for $\mathbf{B}\parallel[110]$. This is consistent with changes in the easy axis of magnetization \cite{SluchankoBogach08, KunimoriKotani11, KunimoriSera12}, which first points along $[110]$ in small fields below the domain-selection transition, then changes to $[001]$ in phase~III above the domain-selection field, then switches briefly to $[111]$ in phase phase~III$^\prime$, and in higher fields changes again to $[110]$ in phase~II. We also observe that for $\mathbf{B}\parallel[110]$, the stability range of phase III$^\prime$ is nearly vanishing, but it increases with La substitution and reaches approximately 1~T in Ce$_{0.7}$La$_{0.3}$B$_6$.

The approximate region occupied with phase~IV is marked with white dashed lines and gray shading in Figs.~\ref{Fig:PhaseDiagrams}\,(c2,\,d2). In Ce$_{0.72}$La$_{0.28}$B$_6$ it is seen as a sharp step in the magnetic torque at low fields [see Fig.~\ref{Fig:PhaseDiagrams}\,(c1)] and nearly coincides with the domain-selection transition of phase~III. The proximity of the two transitions adds some uncertainty to the identification of the corresponding contours in this region of the phase diagram, especially along the $[110]$ field direction. Here thermodynamic measurements by Jang \textit{et al.} done on a piece of the same sample \cite{JangPortnichenko17} show an anomaly in the 0.2~K magnetocaloric sweep, centered around 0.7~T, with a barely resolved two-peak structure. It coincides with an anomaly in the $B$ dependence of the magnetic Bragg intensity at $\left(\frac{1}{2}\frac{1}{2}\frac{1}{2}\right)$. Both anomalies look qualitatively identical to those associated with the domain motion in phase~III at lower La concentrations, where phase~IV is absent~\cite{JangPortnichenko17}. On the other hand, phase IV in Ce$_{0.7}$La$_{0.3}$B$_6$ is suppressed at a somewhat higher field between 0.95 and 0.10~T, judging by the direct polarization-dependent resonant x-ray diffraction measurements \cite{MatsumuraMichimura14} and by the splitting of the $(444)$ structural Bragg peak in the synchrotron diffraction data~\cite{InamiMichimura14}. There is a second anomaly in the $B$ dependence of the $\left(\frac{1}{2}\frac{1}{2}\frac{1}{2}\right)$ magnetic intensity in our neutron data at roughly the same field \cite{JangPortnichenko17}, suggesting that the AFM (phases~III) and AFO (phase~IV) order parameters must coexist in a narrow field range, and that the selection of AFM domains occurs \emph{inside} phase~IV for $\mathbf{B}\parallel[110]$, but \emph{outside} phase~IV for $\mathbf{B}\parallel[001]$. This suggestion requires a direct verification by measurements of the Bragg intensity at $\mathbf{q}_1$ as a function of field along $[110]$.

Finally, in Ce$_{0.5}$La$_{0.5}$B$_6$ the region occupied by phase~IV becomes nearly isotropic with respect to the field rotation \cite{NakamuraEndo03}, whereas phase~III is suppressed for all field directions except in the vicinity of $[001]$ \cite{NakamuraGoto95}, where it still exists in the form of an elongated lobe in Fig.~\ref{Fig:PhaseDiagrams}\,(d2), which is expected to persist up to $x\approx0.7$ judging from previously published resistivity data \cite{KobayashiSera00}. In all other field directions, our torque magnetometry data reveal two phase transitions, which most likely delineate the much reduced stability region of phase III$^\prime$. The data contained in previous publications \cite{HiroiKobayashi97, KobayashiSera00, NakamuraSato03, AkatsuGoto04, JangPortnichenko17, GrunerKim19} were restricted to $\mathbf{B}\parallel[110]$ and $[001]$ and therefore could not reveal the existence of phase III$^\prime$ at $x=0.5$ or higher, which can be clearly seen only around $\mathbf{B}\parallel[111]$ or $[112]$. It would be desirable to verify the magnetic structure of this phase in follow-up works by diffraction measurements.

\vspace{-3pt}\section{\label{Sec:Theory}Theory}\vspace{-2pt}

After we have presented the experimental field-angular magnetic phase diagrams of Ce$_{1-x}$La$_x$B$_6$ and identified the corresponding phases, it is important to understand what essential physics determines their strong anisotropy. Is it really necessary to understand the magnetic structure and microscopic origin of every phase to predict the approximate shape of its stability region? Is the anisotropy consistent with cubic lattice symmetry, or can it serve as evidence for electronic or structural symmetry breaking? To answer these questions, in this section we consider a simple theoretical model consisting of Ce$^{3+}$ ions in the cubic crystal field of the surrounding B$_{24}$ cluster in the form of a truncated cube, which interact by a single nearest-neighbor exchange interaction. By calculating the crystal-field scheme of this model in an applied magnetic field, we obtain the field dependence of magnetization, which can be compared directly with the experimental phase diagrams.

\begin{figure*}[t!]
\centerline{\includegraphics[width=0.9\textwidth]{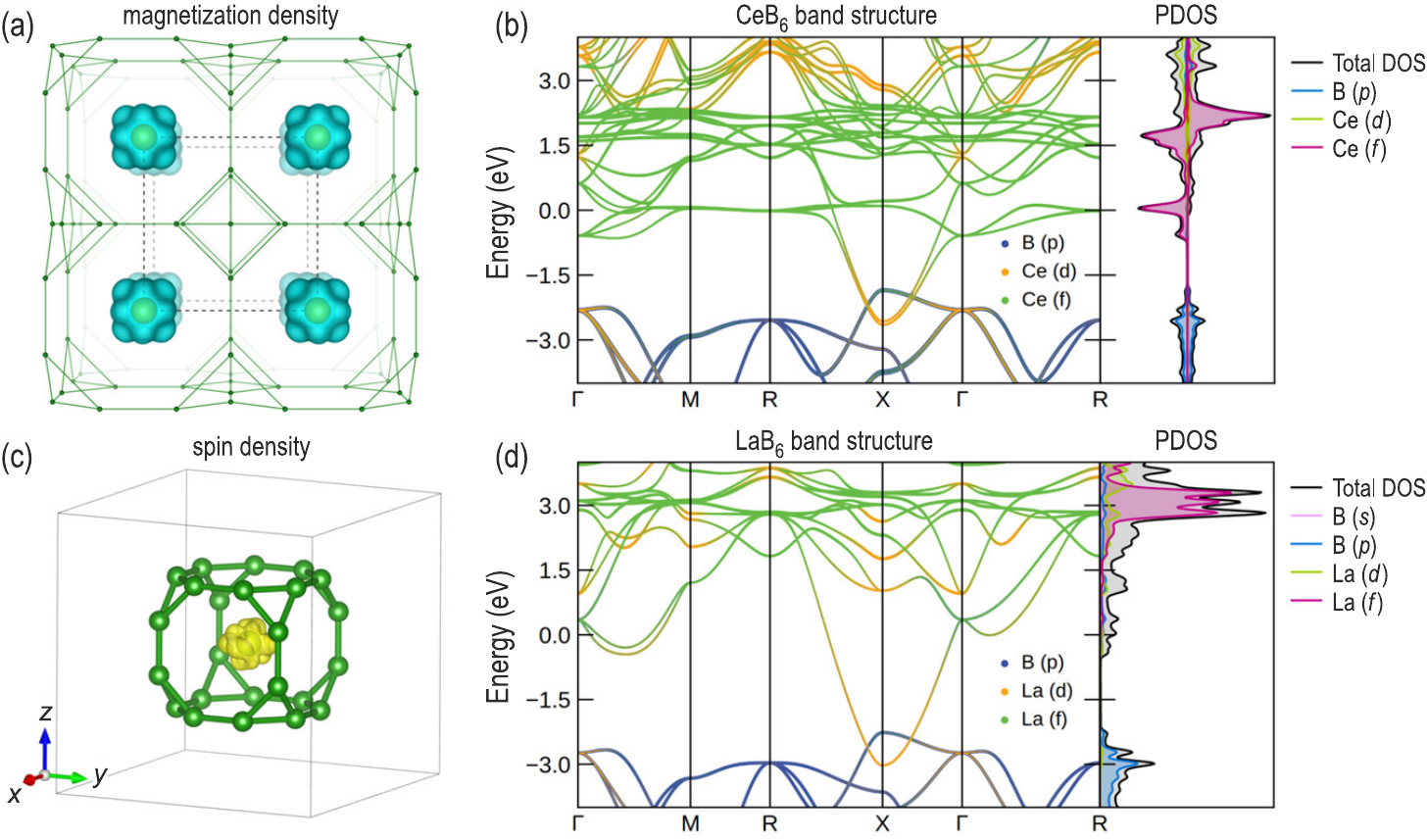}}
\caption{(a)~Isosurfaces of the magnetization density in CeB$_6$ from DFT+$U$/PBE/PAW calculations. (b,\,d)~Band structures and projected densities of states for CeB$_6$ and LaB$_6$ systems. (c)~Natural spin density for the ground-state quartet calculated at the CASSCF level of theory.}
\label{Fig:Theory1}
\end{figure*}

\begin{table*}[t!]
\caption{Decomposition of SOC states of the low-energy multiplet of Ce$^{3+}$ in the $|J,m_J\rangle$ basis for doublets ($D_1$--$D_3$) and for the $\Gamma_8$~quartet\,--\,$\Gamma_7$~doublet ($Q_1$,\,$D_3$), including the corresponding $g$ tensors.}\label{Tab:Theory}
\begin{center}
\begin{tabular}{@{~}c@{~~}|@{~~}cc@{~~}|@{~~}cccccc@{~~}|@{~~}lll@{~}}
\toprule
\textbf{State} & \multicolumn{2}{c@{~~}|@{~~}}{\textbf{SOC energy}} & \textbf{$|\text{\textpm 5/2}\rangle$} & \textbf{$|\text{\textpm 3/2}\rangle$} & \textbf{$|\text{\textpm 1/2}\rangle$} & \textbf{$|\text{\textmp 1/2}\rangle$} & \textbf{$|\text{\textmp 3/2}\rangle$} & \textbf{$|\text{\textmp 5/2}\rangle$} & \multicolumn{1}{c}{\textbf{\textit{g$_\text{x}$}}} & \multicolumn{1}{c}{\textbf{\textit{g$_\text{y}$}}} & \multicolumn{1}{c}{\textbf{\textit{g$_\text{z}$}}}\\
& (cm$^{-1}$) & (meV) &  &  &  &  &  &  &  &  & \\
\midrule
$D_1$ & \multicolumn{2}{c@{~~}|@{~~}}{0.000}    & --- & --- & 0.65 & 0.35 & --- & --- & 2.67 & 2.24 & 0.94 \\
$D_2$ & 0.015 & 0.002                         & 0.48 & 0.07 & --- & --- & 0.10 & 0.34 & 1.18 & 1.61 & 3.01 \\
$D_3$ & 363 & 45.0                            & --- & 0.83 & --- & --- & --- & 0.17 & 1.50 & 1.50 & 1.50 \\ \midrule
$Q_1$ & \multicolumn{2}{c@{~~}|@{~~}}{$\sim$0.0$^\ast$} & --- & --- & --- & --- & --- & --- & 1.00$^\#$ & 1.00$^\#$ & 1.00$^\#$ \\
$D_3$ & 363 & 45.0                            & --- & 0.83 & --- & --- & --- & 0.17 & 1.50 & 1.50 & 1.50 \\
\bottomrule
\multicolumn{12}{l}{$^\ast$Within proposed theoretical framework we observed a tiny gap of 0.015~cm$^{-1\strut} \approx 2~\upmu\text{eV}$.}\\
\multicolumn{12}{l}{$^\#$No twofold degeneracy estimated for effective $\tilde{S} = 3/2$.}
\end{tabular}
\end{center}
\end{table*}

\vspace{-2pt}\subsection{\label{SubSec:TheoryCeB6}\textit{Ab-initio} calculations of the local anisotropy in C\lowercase{e}B$_\text{6}$}

\subsubsection{DFT level}

As the first step, we have modeled equilibrium geometry and Bader charges in CeB$_6$ in the framework of density functional theory (DFT) using \textsc{VASP}~{\tt v}.5 suit (see Appendix~\ref{App:Methods}). Figure~\ref{Fig:Theory1} summarizes the obtained results. The spin-polarized solution predicts a normal metallic ground state [Fig.~\ref{Fig:Theory1}\,(b)] in accordance with the available literature data \cite{HungJeng20}. The band behavior around the Fermi level is dictated by $4f$ orbitals with a single electron, which are also responsible for the isotropic on-site magnetization density plotted in Fig.~\ref{Fig:Theory1}\,(a).

\begin{figure*}[t]
\centerline{\includegraphics[width=0.89\textwidth]{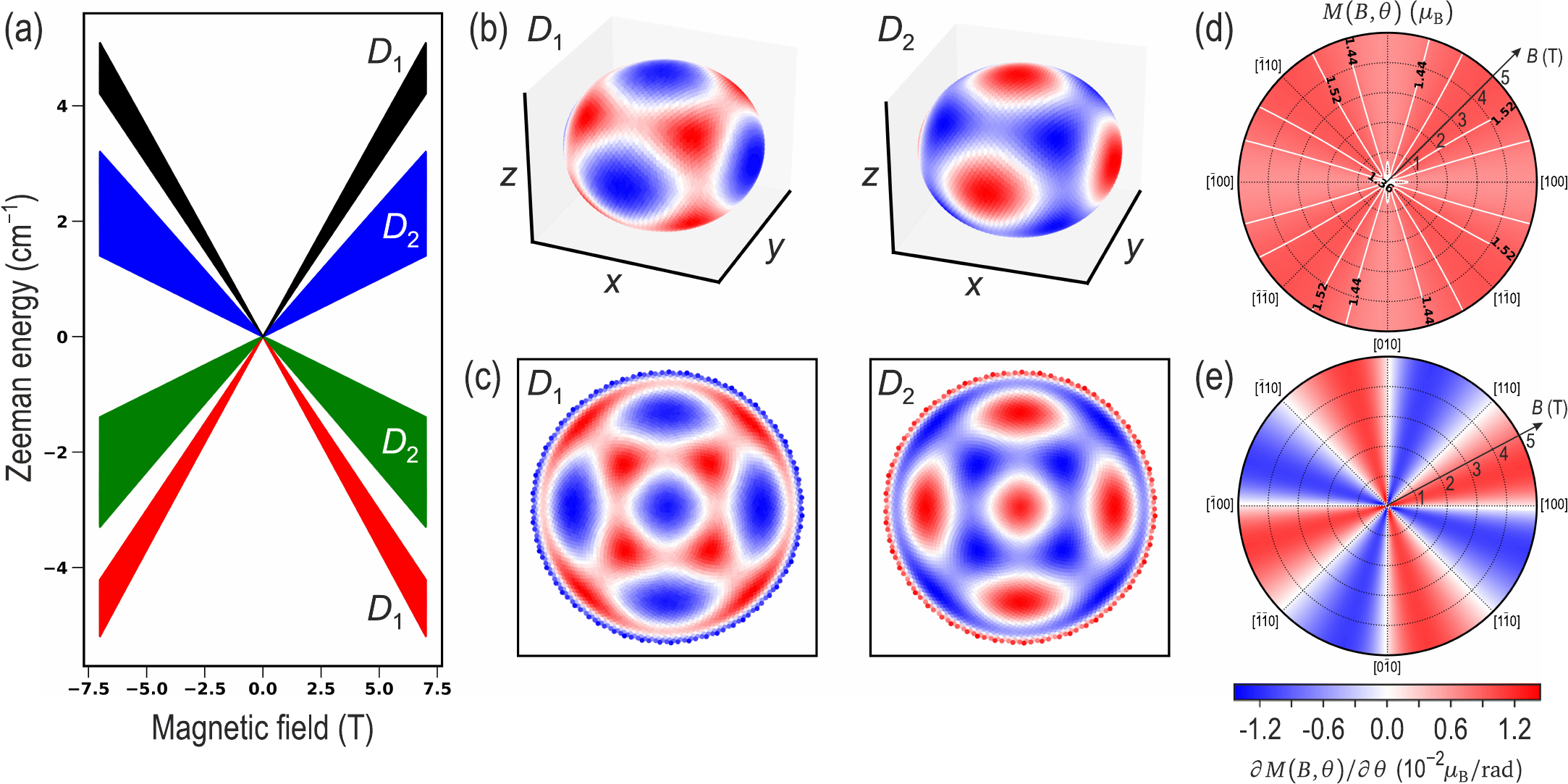}}
\caption{(a)~A schematic Zeeman diagram for the single-ion $\hamcal_\text{m1}$ model in different orientations of $\textbf{B}$ on a spherical grid. The shading covers an energy range of the Zeeman-split states due to the field-angular anisotropy of the $g$ factors. (b,\,c)~Anisotropy of the state magnetization $M_{D_i} \propto \partial E_{D_i}(B)/\partial B|_{B \to 0}$ for the $D_1$ and $D_2$ doublets (see Tables \ref{Tab:Appendix1} and \ref{Tab:Appendix2} in Appendix~\ref{App:Tables}), presented as color maps on a unit sphere and in a polar projection onto the 2D plane, respectively. (d,\,e)~Polar maps of the magnetization, $M(B,\theta) \propto \partial F/\partial B$, and the corresponding magnetic-field derivative of the torque, $\partial\mathbold{\tau}/\partial B \propto \partial^2 F/\partial\theta\partial B \propto \partial M(B,\theta)/\partial\theta$, respectively, as a function of magnetic-field strength $B$ from 0 to 5~T and angle $\theta$ in the $ab$ plane for $\textbf{B} \perp [001]$.}
\label{Fig:Theory2}
\end{figure*}

\subsubsection{CASSCF level}

Although the mean-field DFT+$U$ result correctly predicts the $4f^1$ configuration of CeB$_6$, the exact $4f$ multiplet (Ce$^\text{III}$, $^{2\!}$\textit{F}$_{\!\text{5/2}}$) structure cannot be recovered. This information is crucial for local and extended anisotropy modeling. To have a proper model for multiplet structures and CEF parameters, we performed multi-configurational complete active space self-consistent field (CASSCF) quantum chemistry computations using \textsc{OpenMolcas} code \cite{KresseHafner93}. Additional details of the methods used in our \textit{ab initio} calculations can be found in Appendix~\ref{App:Methods}.

We used a [CeB$_{24}$]$^{3+}$ cluster model with all boron atoms at the vertices of the truncated cube substituted by point charges of $-1.5e$. The charge value is picked empirically, so that the energy difference between the ground-state $\Gamma_8$ quartet and the excited $\Gamma_7$ doublet matches the energy of the $\Gamma_8 \rightarrow \Gamma_7$ CEF transition ($\sim$46~meV or 372~cm$^{-1}$ at room temperature) observed with inelastic neutron scattering and Raman spectroscopy \cite{ZirngieblHillebrands84, ZirngieblHillebrands85, SatoKunii84, LoewenhauptCarpenter85, YeKung19}. The obtained \textit{ab initio} spin-orbit coupling (SOC) states were analyzed in terms of projection on the lowest-energy atomic multiplet ($J=\text{5/2}$). Furthermore, the \textit{ab initio} CEF parameters $B_k^{\,q}$ (in Stevens-operator notation \cite{Rudowicz85}) were derived for the model Hamiltonian design (see Appendix~\ref{App:Tables}).

The decomposition of SOC states in the $|J,m_J\rangle$ basis and the $g$-tensor structure for the lowest-energy atomic multiplet are summarized in Table~\ref{Tab:Theory}. The ground-state quartet is composed of two doublets, $D_1$ and $D_2$, that nearly coincide in energy (separated by a negligibly small energy gap of 0.015~cm$^{-1}$) and are composed in equal measure of $|\pmmp\text{1/2}\rangle$ and $|\pmmp\text{5/2}\rangle$ projections of the total angular momentum $\mathbf{J}$. This ground-state quartet $Q_1$ is characterized by an isotropic natural spin density plotted in Fig.~\ref{Fig:Theory1}\,(c).

Using the derived \textit{ab initio} CEF parameters $B^{\,q}_k$ listed in Table~\ref{Tab:Appendix2} in Appendix~\ref{App:Tables}, we can write down the following model Hamiltonian for a single Ce$^{3+}$ ion in the CeB$_6$ system:
\begin{equation}\label{Eq:Ham1}
\hamcal_\text{m1}=\hamcal_{\text{CF}_\text{Ce1}} + \hamcal_\text{Zee}=\kern-3pt\mathlarger{\sum}_{k=2,4,6}\,\sum_{q=-k}^k{\kern-3pt}B^{\,q}_k\hat{O}^{\,q}_k + \mu_\text{B}\hat{g}\cdot\hat{J\kern2.3pt}\kern-3pt\cdot\mathbf{B}
\end{equation}
in terms of the Stevens operators $\hat{O}^{\,q}_k$~\cite{Rudowicz85}. After including exchange interactions $\mathcal{j}_{12}$ between the first-nearest-neighbor Ce$^{3+}$ ions, the Hamiltonian takes the form~\cite{Rotter04}
\begin{equation}\label{Eq:Ham2}
\hamcal_\text{m2}=\hamcal_{\text{CF}_\text{Ce1}} + \hamcal_{\text{CF}_\text{Ce2}} + \hamcal_\text{Zee} -2\mathcal{j}_{12\,}^{\phantom{~}}{\hat{J\kern2.3pt}\kern-3pt}_{\text{Ce1}}{\hat{J\kern2.3pt}\kern-3pt}_{\text{Ce2}}.
\end{equation}
Here $\hamcal_{\text{CF}_\text{Ce1,2}}$ are the single-ion CF Hamiltonians for the two Ce sites, described by the corresponding \textit{ab initio} crystal field parameters $B_k^{\,q}$ in Stevens notation, and $\hamcal_\text{Zee}$ is the Zeeman energy, both defined according to Eq.~\eqref{Eq:Ham1}. The additional term $-2\mathcal{j}_{12\,}^{\phantom{~}}{\hat{J\kern2.3pt}\kern-3pt}_{\text{Ce1}}{\hat{J\kern2.3pt}\kern-3pt}_{\text{Ce2}}$ stands for the exchange energy between nearest-neighbor localized lanthanide spins ${\hat{J\kern2.3pt}\kern-3pt}_{\rm Ce1}$ and ${\hat{J\kern2.3pt}\kern-3pt}_{\rm Ce2}$.

The model given by Eq.~\eqref{Eq:Ham1} helps to uncover the spatial structure of the single-ion magnetic anisotropy, which is shown in Fig.~\ref{Fig:Theory2}. Expectedly, the anisotropy of the state magnetization $M_{D_i} \propto \partial E_{D_i}(B)/\partial B|_{B \to 0}$ for the $D_1$ and $D_2$ doublets in Figs.~\ref{Fig:Theory2}\,(b,\,c) inherits the shape of the natural spin density for the ground-state quartet in Fig.~\ref{Fig:Theory1}\,(c). The total magnetization $M(B,\theta)$, plotted in Fig.~\ref{Fig:Theory2}\,(d) for $\mathbf{B} \perp [001]$ as a function of the magnitude and direction of the magnetic field in the $ab$ plane, has a four-fold symmetry with minima in magnetization along the $\langle 100 \rangle$ and maxima along the $\langle 110 \rangle$ directions in the crystal. The magnetic-field derivative of the torque, $\partial\mathbold{\tau}(B,\theta)/\partial B$, which can be also calculated by differentiating $M(B,\theta)$ with respect to the field angle $\theta$, acquires the sign-changing $\sin\theta\cos\theta$ angular dependence presented in Fig.~\ref{Fig:Theory2}\,(e). In the single-ion case, the field dependence of this quantity is trivial.

\subsubsection{Two-site model}\vspace{-2.5pt}

\begin{figure}[t]
\includegraphics[width=\columnwidth]{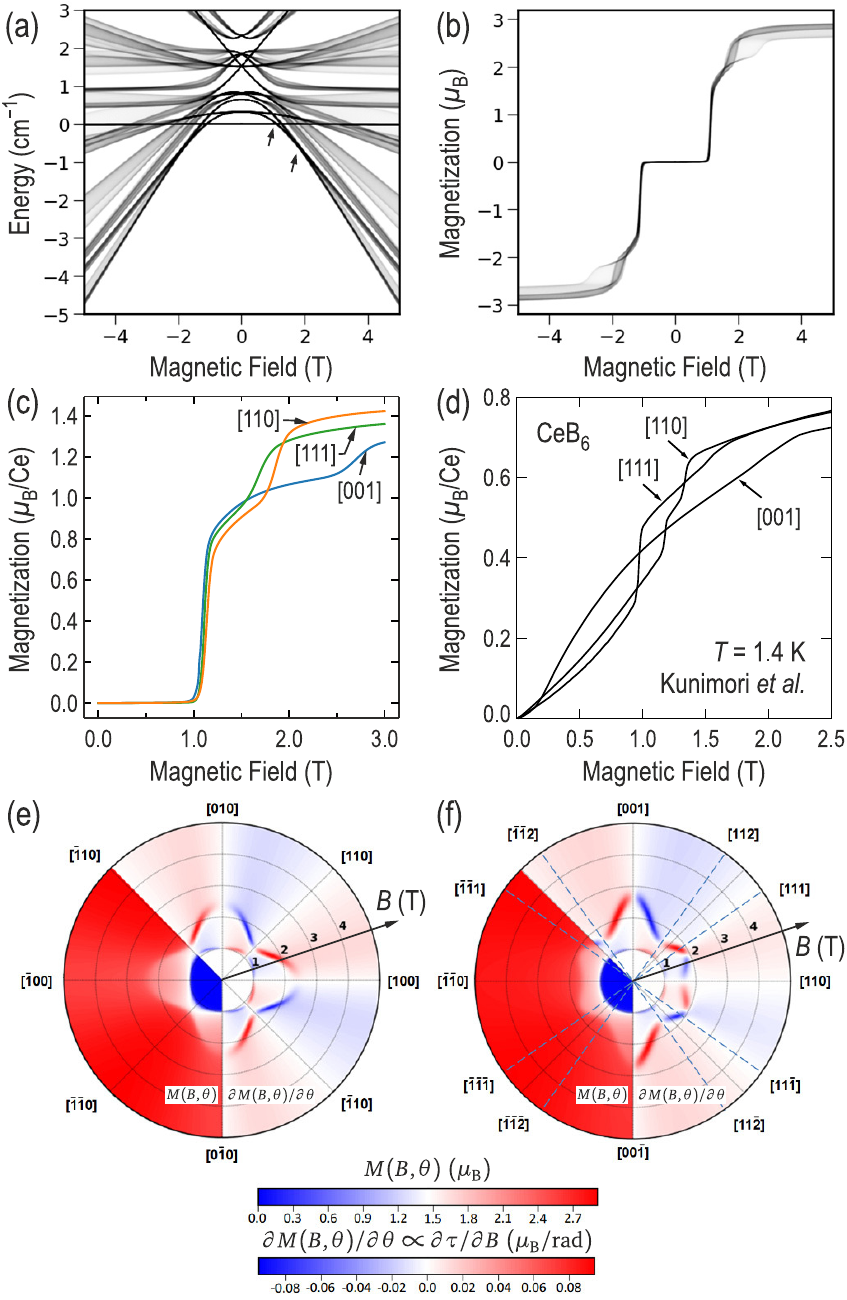}
\caption{(a)~Superposed Zeeman diagrams and (b)~a set of magnetization curves obtained in the two-site model ($\hamcal_\text{m2}$) for magnetic-field scans up to 5~T along different directions in the $\textbf{B}\perp[1\overline{1}0]$ plane. The black arrows in panel (a) indicate ground-state level crossings responsible for the appearance of magnetization steps. (c)~Calculated magnetic-field dependence of the total magnetization in the two-site model ($\hamcal_\text{m2}$) for the three high-symmetry directions in the cubic crystal. (d)~Experimental magnetization curves, reproduced with permission from Ref.~\citenum{KunimoriKotani11} for comparison. (e,\,f)~The total magnetization $M(B,\theta)$ and its angle derivative $\partial M(B,\theta)/\partial\theta$, which is equivalent to the experimentally measured magnetic-field derivative of the torque $\partial\mathbold{\tau}/\partial B$, presented as polar maps as a function of magnetic-field strength $B$ from 0 to 5~T and its in-plane direction $\theta$ for $\textbf{B} \perp [001]$ and $\textbf{B} \perp [1\overline{1}0]$, respectively.}
\label{Fig:Theory3}
\end{figure}

As the next step, we consider two interacting nearest-neighbor Ce$^{3+}$ sites described by Eq.~\eqref{Eq:Ham2}. The results of the corresponding calculation are presented in Fig.~\ref{Fig:Theory3}. The combination of ground-state quartets on the two Ce$^{3+}$ ions results in 16 eigenstates of the Hamiltonian $\hamcal_{m1}$, and a heavily crowded Zeeman diagram shown in Fig.~\ref{Fig:Theory3}\,(a) emerges in an applied magnetic field. Here we superposed multiple orientations of $\mathbf{B}$ to visualize the ranges of eigenstate energies that would result as the field is rotated with respect to the crystal. This Zeeman diagram reveals two level crossings (indicated with black arrows), which represent transitions between different quantum-state compositions of the local ground state, starting with the original AFM configuration below 1~T. They occur between 1 and 3~T, where phase transitions between different long-range-ordered magnetic ground states are observed experimentally (Sec.~\ref{Sec:Experiment}). Similarly to the single-ion system, magnetization $M(B,\theta) \propto \partial F/\partial B$ can be calculated, which now acquires a nontrivial field dependence with several steps, as shown in Fig.~\ref{Fig:Theory3}\,(b), that appear whenever the composition of the ground state changes due to a level crossing.

The calculated field dependence of the magnetization for three high-symmetry directions of the magnetic field are compared in Fig.~\ref{Fig:Theory3}\,(c). The magnetization curves are qualitatively similar to the experimental ones, which we reproduce in Fig.~\ref{Fig:Theory3}\,(d) from Kunimori \textit{et al}~\cite{KunimoriKotani11}. Apparently, the two-site model correctly reproduces the change in the easy axis of the magnetization from $[001]$ at low fields through $[111]$ at intermediate fields to [110] at high fields. This is remarkable, because the magnetization steps in the experimental data correspond to transitions between long-range-ordered phases~III, III$^\prime$, and II, yet the magnetic unit cells of all these three phases are much larger than the size of our two-site cluster. It is therefore clear that our simple theoretical model is unable to describe the structure of magnetically ordered phases, and still it can reproduce the transition fields between them and the corresponding anisotropic behavior of the magnetization at least qualitatively.

The magnetic-field derivative of the torque, which is measured in experiment, can be calculated from the angle dependence of the magnetization according to Eq.~\eqref{Eq:Derivatives}. In Figs.~\ref{Fig:Theory3}\,(e,\,f), we plot both magnetization $M$ and its angle derivative, $\partial M(B,\theta)/\partial\theta$, as a function of magnetic field rotated in the plane orthogonal to $[001]$ and $[110]$, respectively. These polar plots bear a striking similarity to the corresponding experimental maps in Figs.~\ref{Fig:PhaseDiagrams} and \ref{Fig:CeB6_HK0}, qualitatively reproducing not only the magnitude of the anisotropy (approximately a factor of 2 in both cases), but even the shape of the resulting transitions in field space. Both the extended lobe protruding in the $\langle001\rangle$ directions and the local minima along the $\langle110\rangle$ and $\langle111\rangle$ directions with a small local maximum between them are captured by the calculation. However, our theoretical model underestimates the number of transition, which is not surprising in view of the small size of the cluster with only two Ce$^{3+}$ ions.

In fact, the level of agreement between experiment and theory is strikingly good for our oversimplified model. Indeed, the experimental features observed in torque magnetometry correspond to phase transitions between long-range-ordered AFM or AFQ states. Not only is a local two-site model unable to realize long-range order, but even the sizes of magnetic unit cells in phases II, III, and III$^\prime$ are considerably larger than two atoms. For instance, the zero-field ground state of CeB$_6$ (phase III) is known to represent a double-$\mathbf{q}$ AFM order characterized by a pair of propagation vectors, $\mathbf{q}_1=(\frac{1}{4}\frac{1}{4}0)$ and $\mathbf{q}_2=(\frac{1}{4}\frac{1}{4}\frac{1}{2})$, which implies a magnetic unit cell with $4\times4\times2=32$ magnetic ions \cite{EffantinRossat-Mignod85, ZaharkoFischer03}. An \textit{ab initio} quantum-chemistry calculation of such a cluster is unfeasible, as it would have to consider $4^{32} \approx 1.8\times10^{19}$ eigenstates of the full $\hamcal_\text{m16}$ Hamiltonian that could be constructed by extending Eq.~\eqref{Eq:Ham2} by analogy to a cluster of 32 magnetic atoms. If such a calculation was possible, we expect that it would do a better job of distinguishing different magnetic ground-state configurations and in reproducing the sequence of metamagnetic phase transitions observed in experiments more accurately.

An inevitable conclusion from these considerations is that the anisotropy and shape of phase boundaries in field-angular space has a much more fundamental origin than the ordered states themselves. While they cannot be captured in the single-ion model, inclusion of just a single interaction with the nearest neighbors turns out sufficient for a qualitative description of the main features in the field-angular phase diagram.

\section{\label{Sec:Summary}Summary and conclusions}\vspace{-2pt}

In summary, we have investigated experimentally the evolution of the field-angular phase diagrams of cerium hexaboride and its Ce$_{1-x}$La$_x$B$_6$ solid solutions with increasing La concentration, $0 \leq x \leq 0.5$, which were extensively studied previously but only for a few selected high-symmetry field directions. Our results show excellent agreement with most of the previously published studies on the magnetic properties of pure and La-doped CeB$_6$ but in addition reveal some important details of the phase diagrams that only become clear due to the continuous field rotation. In particular, we demonstrate that the single-$\mathbf{q}$ AFM phase III$^\prime$ persists at least up to 50\% La doping level for field directions $\langle112\rangle$ and $\langle111\rangle$, whereas the double-$\mathbf{q}$ phase~III exists only in narrow lobes surrounding the $\mathbf{B}\parallel[001]$ and equivalent field directions.

We have also demonstrated that a local model consisting of two Ce$^{3+}$ ions in the cubic crystal field with a single effective interaction between them can qualitatively capture the anisotropic features of the field-angular phase diagram of cerium hexaboride without violating the underlying cubic symmetry of the lattice. Clearly, such a primitive model is insufficient to describe the ordered states in Ce$_{1-x}$La$_x$B$_6$, yet it can capture transitions between different local CEF ground states stabilized in magnetic field that ultimately participate in magnetic ordering. Consequently, under the assumption that the effective interactions between different degrees of freedom on neighboring ions are field-independent, stability ranges of different ordered phases tend to inherit those of the local ground states, notwithstanding the microscopic structure of these phases that can depend on further-neighbor interactions and other details not captured by the local model. We therefore conclude that the anisotropy of phase boundaries separating long-range-ordered magnetic phases in the field-angular phase diagram of CeB$_6$ are of much more fundamental origin than these phases themselves and are therefore much easier to describe. Remarkably, even an oversimplified purely local two-site model that does not break the cubic lattice symmetry and neglects the metallic character of the compound appears sufficient to qualitatively capture the anisotropic shape of the phase stability regions in the field-angular phase diagram.\vspace{-2pt}

\begin{acknowledgments}
We thank N.~Yu.~Shitsevalova and V.\,B.~Filipov from the I.~M. Frantsevich Institute for Problems of Materials Science of NAS, Kyiv, Ukraine, for providing the single-crystal samples for our torque magnetometry measurements and S.~Nikitin for preliminary magnetization measurements that preceded the present work. This work is supported, in part, by the German Research Foundation (DFG) under individual research grants \mbox{IN~209/3-2}, \mbox{PO~2621/1-1}, \mbox{AV~169/3-1}, and by the W\"urzburg-Dresden Cluster of Excellence on Complexity and Topology in Quantum Matter\,---\,\textit{ct.qmat} (EXC 2147, project-id 390858490). The experimental part of this work was performed at the National High Magnetic Field Laboratory, which is supported by the National Science Foundation Cooperative Agreement No.~DMR-1644779 and the state of Florida.
\end{acknowledgments}

\appendix

\onecolumngrid

\section{Theoretical methods and models}\label{App:Methods}

\twocolumngrid


\textbf{\small\textsf{DFT modeling.}} All structures at the \mbox{DFT+$U$/PBE/PAW} level of theory were optimized using projector augmented-wave method as implemented in the \textsc{VASP}~{\tt v}.5 code and the standard pseudopotential~\cite{KresseHafner93, PerdewErnzerhof96, KresseJoubert99}.

\textbf{\small\textsf{\textit{Ab initio} modeling}}. The first-principles CASSCF calculations were done at the DKH2/CAS(1,7)/RASSI-SO/VDZ-RCC level of theory using \textsc{OpenMolcas}~\cite{RoosLindh08, AquilanteAutschbach20}. In the complete active space (CAS) approach, the total spin of the 4\textit{f} shell with one electron is $S=\text{1/2}$. Having solved the spin-free CAS problem, all seven roots were used in further state interaction modeling using the SOC Hamiltonian. Finally, for all doubly degenerate SOC states the $g$ tensors were computed using first-order perturbation theory. SOC-states decomposition in $J=\text{5/2}$ multiplet and the \textit{ab initio} CEF scheme were derived using the \textsc{Single\_Aniso} module~\cite{ChibotaruUngur12}.

\textbf{\small\textsf{Model Hamiltonians}}. The model Hamiltonian was solved, and the magnetic properties of the system were analyzed using \textsc{PHI} code and in-house Python scripts~\cite{ChiltonAnderson13}. The generalized model Hamiltonian for the extended system model was solved using \textsc{McPhase} libraries~\cite{Rotter04}.

\onecolumngrid

\section{Calculated parameters of the CEF state for CeB$_\text{6}$}\vspace{-4pt}
\label{App:Tables}

Tables \ref{Tab:Appendix1} and \ref{Tab:Appendix2} summarize the results of \textit{ab initio} modeling including energies and Stevens parameters of all SOC states for the $J=5/2$ multiplet. 

\twocolumngrid

\begin{table}[t!]
\caption{Parameters of the SOC states for the [CeB$_{24}$]$^{3+}$ cluster at the DKH2/CAS(1,7)/RASSI-SO/VDZ-RCC level of theory.}\label{Tab:Appendix1}
\begin{center}
\begin{tabular}{l@{~~~}c@{\qquad}r@{\qquad}c}
\toprule
\textbf{SOC-id} & \textbf{Energy (eV)} & \multicolumn{1}{c}{\textbf{Energy (cm$^\text{--1}$)}} & \textbf{\textit{J}} \\
\midrule
{\tt ~1} & {\tt 0.0000000000} & {\tt 0.0000}~~~ & {\tt 2.5} \\
{\tt ~2} & {\tt 0.0000000000} & {\tt 0.0000}~~~ & {\tt 2.5} \\
{\tt ~3} & {\tt 0.0000018402} & {\tt 0.0148}~~~ & {\tt 2.5} \\
{\tt ~4} & {\tt 0.0000018402} & {\tt 0.0148}~~~ & {\tt 2.5} \\
{\tt ~5} & {\tt 0.0449916213} & {\tt 362.8819}~~~ & {\tt 2.5} \\
{\tt ~6} & {\tt 0.0449916213} & {\tt 362.8819}~~~ & {\tt 2.5} \\
{\tt ~7} & {\tt 0.2856616206} & {\tt 2304.0164}~~~ & {\tt 3.5} \\
{\tt ~8} & {\tt 0.2856616206} & {\tt 2304.0164}~~~ & {\tt 3.5} \\
{\tt ~9} & {\tt 0.3146648966} & {\tt 2537.9436}~~~ & {\tt 3.5} \\
{\tt 10} & {\tt 0.3146648966} & {\tt 2537.9436}~~~ & {\tt 3.5} \\
{\tt 11} & {\tt 0.3146672418} & {\tt 2537.9625}~~~ & {\tt 3.5} \\
{\tt 12} & {\tt 0.3146672418} & {\tt 2537.9625}~~~ & {\tt 3.5} \\
{\tt 13} & {\tt 0.3571040930} & {\tt 2880.2388}~~~ & {\tt 3.5} \\
{\tt 14} & {\tt 0.3571040930} & {\tt 2880.2388}~~~ & {\tt 3.5} \\
\bottomrule
\end{tabular}
\end{center}
\end{table}

~

\vfill\eject

\begin{table}[t!]
\caption {$B_k^{\,q}$ parameters in Stevens-operator notation for the crystal-field Hamiltonian of the $J=\text{5/2}$ multiplet.} \label{Tab:Appendix2}
\begin{center}
\begin{tabular}{l@{\quad}r@{\quad}r}
\toprule
\textbf{\textit{k}} & \textbf{\textit{q}} & \multicolumn{1}{c}{\textbf{\textit{B}$_\text{\textit{k}}^\text{\,\textit{q}}$ (cm$^\text{--1}$)}} \\
\midrule
{\tt 2} & {\tt -2} & {\tt -0.32372802917775E-05} \\
{\tt 2} & {\tt -1} & {\tt -0.10491200943645E-04} \\
{\tt 2} & {\tt  0} & {\tt  0.60504190785733E-01} \\
{\tt 2} & {\tt  1} & {\tt -0.55654827296684E-02} \\
{\tt 2} & {\tt  2} & {\tt -0.14444563753813E-02} \\
{\tt 4} & {\tt -4} & {\tt  0.73574190082391E-07} \\
{\tt 4} & {\tt -3} & {\tt  0.18700751961646E-06} \\
{\tt 4} & {\tt -2} & {\tt  0.31185896445511E-07} \\
{\tt 4} & {\tt -1} & {\tt  0.63451704654471E-07} \\
{\tt 4} & {\tt  0} & {\tt -0.10039210366818E+01} \\
{\tt 4} & {\tt  1} & {\tt -0.41719162424810E-04} \\
{\tt 4} & {\tt  2} & {\tt  0.25009971775102E-04} \\
{\tt 4} & {\tt  3} & {\tt -0.13176446112330E-03} \\
{\tt 4} & {\tt  4} & {\tt -0.50681897587772E+01} \\
\bottomrule
\end{tabular}
\end{center}
\end{table}

\onecolumngrid\vspace*{-3.5em}\twocolumngrid

\bibliographystyle{my-apsrev}

\bibliography{CeLaB6_FieldAngle}\vspace{-3pt}
\onecolumngrid

\end{document}